\def\Msun{\hbox{M$_\odot$}}
\def\Msol{\hbox{M$_\odot$}}
\def\Zsol{\hbox{Z$_\odot$}}

\def\kms{\hbox{km$\,$s$^{-1}$}}
\def\cmt{\hbox{cm$^{-3}$}}

\def\one{\,{\sc i}}             % for producing Na I as Na\one\ etc.
\def\two{\,{\sc ii}}
\def\three{\,{\sc iii}}
\def\four{\,{\sc iv}}
\def\five{\,{\sc v}}

\def\ha{H$\alpha$}

\documentclass[iop,apj]{emulateapj}

\usepackage{amsmath}
\usepackage{amssymb}
\usepackage{mathrsfs}
\usepackage[usenames,dvipsnames]{color,colortbl}
\usepackage{soul} %text highlighting (has to be used with the color package to work)
\usepackage{upgreek}
\usepackage{graphicx}
\usepackage{overpic}
\usepackage{threeparttable}
\shorttitle{Very Massive Stars in NGC 5253}
\shortauthors{L.\ J.\ Smith et al.}

\begin{document}

%%%%%%%%%%%%%%%%%%%%%%%%%%%%%%%%%%%%%%
%AUTHORS
%%%%%%%%%%%%%%%%%%%%%%%%%%%%%%%%%%%%%%

\title{The Very Massive Star Content of the Nuclear Star Clusters in NGC 5253}
\author{L.\ J.\ Smith\altaffilmark{1}}
\author{P.\ A.\ Crowther\altaffilmark{2}}
\author{D. Calzetti\altaffilmark{3}}
\author{F.\ Sidoli\altaffilmark{4}}

\altaffiltext{1}{Space Telescope Science Institute and European Space Agency, 3700 San Martin Drive, Baltimore, MD 21218; lsmith@stsci.edu}
\altaffiltext{2}{Department of Physics and Astronomy, University of Sheffield, Sheffield S3 7RH, UK}
\altaffiltext{3}{Dept. of Astronomy, University of Massachusetts -- Amherst, Amherst, MA 01003}
\altaffiltext{4}{London Centre for Nanotechnology, University College London, London WC1E 6BT, UK}

\begin{abstract}
The blue compact dwarf galaxy NGC 5253 hosts a very young starburst containing twin nuclear star clusters, separated by a projected distance of 5 pc. One cluster (\#5) coincides with the peak of the H$\alpha$ emission and the other (\#11) with a massive ultracompact H\two\ region. A recent analysis of these clusters shows that they have a photometric age of $1\pm1$~Myr, in apparent contradiction with the age of 3--5~Myr inferred from the presence of Wolf-Rayet features in the cluster \#5 spectrum. We examine {\it Hubble Space Telescope} ultraviolet and {\it Very Large Telescope} optical spectroscopy of \#5 and show that the stellar features arise from very massive stars (VMS), with masses greater than 100~\Msun, at an age of 1--2 Myr.  We further show that the very high ionizing flux from the nuclear clusters can only be explained if VMS are present. We investigate the origin of the observed nitrogen enrichment in the circum-cluster ionized gas and find that the excess N can be produced by massive rotating stars within the first 1~Myr. We find similarities between the NGC 5253 cluster spectrum and those of metal poor, high redshift galaxies. We discuss the presence of VMS in young, star-forming galaxies at high redshift; these should be detected in rest frame UV spectra to be obtained with the James Webb Space Telescope. We emphasize that population synthesis models with upper mass cut-offs greater than 100~\Msun\ are crucial for future studies of young massive star clusters at all redshifts.
\end{abstract}

\keywords{galaxies: dwarf -- galaxies: individual (NGC\,5253) -- galaxies: starburst -- galaxies: star clusters: general -- stars: massive -- stars: Wolf-Rayet}

\newpage

%%%%%%%%%%%%%%%%%%%%%%%%%%%%%%%%%%%%
\section{Introduction} \label{intro}
The blue compact dwarf galaxy NGC 5253 hosts a very young starburst at its center. The near-absence of non-thermal radio emission \citep{beck96} shows that the starburst is too young to have produced many supernovae. 
Wolf-Rayet (W-R) emission features from WN and WC stars were first detected by \citet{Campbell86} and \citet{schaerer97b} in groups of young star clusters, suggesting that the burst is 3--5 Myr old. The triggering process for the current starburst is not clear. An encounter with M83 has been suggested \citep{vdb80,caldwell89} or Cen A \citep{karachentsev07, tully15}. The distance to NGC~5253 is uncertain with values of 3.1--4.0~Mpc quoted in the literature. Here, we adopt a distance of 3.15 Mpc for NGC 5253 \citep{freedman01,davidge07} based on Cepheid and Tip of the Red Giant Branch (TRGB) measurements. We note that the most recent TRGB measurement places NGC~5253 at 3.55~Mpc  \citep{tully13}.
\citet{lopez12} have performed an H\one\ line and 20-cm radio continuum study of NGC 5253. They find that the H\one\ morphology is very disturbed and suggest that the starburst is being triggered by the infall of a diffuse, low metallicity H\one\ cloud along the minor axis. 
The metallicity of NGC 5253 is fairly low with the recent study of \citet{monreal12} giving $12+\log\,$O/H$=8.26$ or 35\% solar  \citep[using the solar oxygen abundance of][]{asplund09}. Recently, \citet{turner15} presented Submillimeter Array images of NGC 5253 showing evidence for a CO streamer falling into the center of the galaxy, and possibly fuelling the current star formation.

\citet{turner98} found that radio maps of the central region are dominated by a single unresolved source with the characteristics of a massive ultracompact H\two\ region. \citet{turner00} resolved this source and find a bright, very dense radio nebula of dimensions $0''.10 \times 0''.04$ ($1.5 \times 0.6$ pc) in the central starburst, which they term the ``supernebula''. The Lyman continuum rate required to excite the nebula is equivalent to a few thousand O7\,V stars within the central $2''$ \citep{crowther99, turner04}.
Hubble Space Telescope (HST) observations with the Near Infrared Camera and Multi Object Spectrometer (NICMOS)
by \citet{alonso04} revealed the presence of a double nuclear star cluster with the two components separated by $0''.3$--$0.''4$ ($\approx 5$ pc). The eastern cluster coincides with the peak of the \ha\ emission in NGC 5253 and is the youngest optical star cluster identified by \citet{calzetti97}. The western cluster is very reddened and is coincident with the supernebula of \citet{turner00}. 

%ref to supernebula is Turner et al. 1998, 2000, Turner & Beck 2004. 
%1200 O7 stars are required to excite 1 pc-sized core of the H II region.
% and 2000 O7 stars within central 5 pc
% and 7000 O7 stars within central 20 pc radius. (=1.2 arcsec)
% PAC - ISO obtains 1000 O7V stars in central 2 arcsec and 2500 in 20-arcsec region.

In a recent paper, \citet{calzetti15} present a detailed analysis of the two nuclear clusters (\#5 and \#11 in their terminology) using HST imaging from the far ultraviolet to the infra-red. By fitting their spectral energy distributions (SED), they find that the two clusters are extremely young with ages of only $1\pm 1$~Myr. The SED for cluster \#5 can be fit with a foreground dust model while a combination of a homogeneous dust-star mixture and a foreground dust screen is required for cluster \#11. The derived cluster masses are $7.5 \times 10^4$ and $2.5 \times 10^5$~\Msun\ for \#5 and \#11 respectively. 
The predicted \ha\ luminosity for cluster \#5 is higher than the observed, attentuation-corrected value, and this discrepancy is interpreted as being due to the leakage of 25--50\% of the ionizing photons into a larger ionized, diffuse region surrounding both clusters, and coincident with the radio nebula. For cluster \#11, the predicted and observed attentuation-corrected \ha\ luminosities agree well and argue in favor of the SED fitting results. The young ages of the two nuclear star clusters are in apparent contradiction with the ages of 3--5 Myr inferred from the presence of W-R stars in cluster \#5 \citep[e.g.][]{monreal10,turner15}. In this paper, we re-examine this aspect by considering whether the W-R features arise from hydrogen-rich very massive stars (VMS; masses $> 100$~\Msun).

\citet{crowther10} found that four stars in R136, the central ionizing cluster of the 30 Doradus region in the Large Magellanic Cloud (LMC), have masses that exceeded the then standard upper mass limit of 150 \Msun\ \citep{figer05}. The four stars they find have initial masses between 165 and 320 \Msun,  based on model fitting.
More recently, \citet{crowther16} have presented a stellar census of the content of R136 from HST/STIS UV spectroscopy. %for 17 adjacent pointings. 
They derive a cluster age of $1.5\pm0.5$~Myr and find that the He\two\ $\lambda1640$ emission flux in the R136 cluster originates exclusively from stars with masses above 100~\Msun.

VMS are also expected to make a significant contribution to the ionizing flux output and account for $\sim$ 25\% of the ionizing photon luminosity in R136  \citep{doran13}. In NGC 5253, clusters \#5 and \#11 contribute about 40--50\% of the total ionization of the galaxy, with the majority coming from cluster \#11. However, 50\% of the ionizing photon rate is unaccounted for in the region of the supernebula \citep{calzetti15}. \citet{turner15} find that they need to invoke a top-heavy initial mass function (IMF) for the two nuclear clusters to provide sufficient Lyman ionizing photons given the age constraint of 4 Myr and an upper limit to the combined cluster mass (derived from the dynamical mass of the gas in the region).

Another remarkable feature of the starburst region in NGC 5253 is the presence of nitrogen-enriched (by a factor of 2--3) nebular material \citep{walsh87, kobulnicky97, lopez07}. \citet{monreal10} have mapped the area of nitrogen enrichment and find it peaks in the giant H\two\ region, which contains the two nuclear star clusters, and extends up to distances of $3''.3$ (50 pc). The source of the nitrogen enrichment is not clear because if it is due to the chemically-enriched winds of W-R stars, then helium should also be enriched, but this has not been conclusively observed \citep{monreal13}. 

Instead, the idea has been put forward \citep{kobulnicky97} that the N enrichment is due to the winds of late O stars where CNO processing can produce N-rich material without the accompanying He enrichment. Since that paper, the effect of rotational mixing in O stars has shown that most fast rotating O stars will show N enrichment at their surfaces. Recently,  rotating evolutionary tracks for VMS with LMC composition have been computed by \citet{yusof13} and \citet{kohler15}. These studies find that the degree of N and He enrichment depends on rotational mixing and mass loss.
In this paper, we explore whether VMS and/or fast-rotating stars can explain the N enrichment associated with the nuclear clusters in NGC 5253. 

This paper is organized as follows. In Section 2, we describe the spectroscopic observations and interpret them in Section 3. We discuss our findings that VMS are present in cluster \#5 in Section 4, and present our summary and conclusions in Section 5.

\begin{figure}
\includegraphics[width=0.45\textwidth,angle=0,clip=true]
{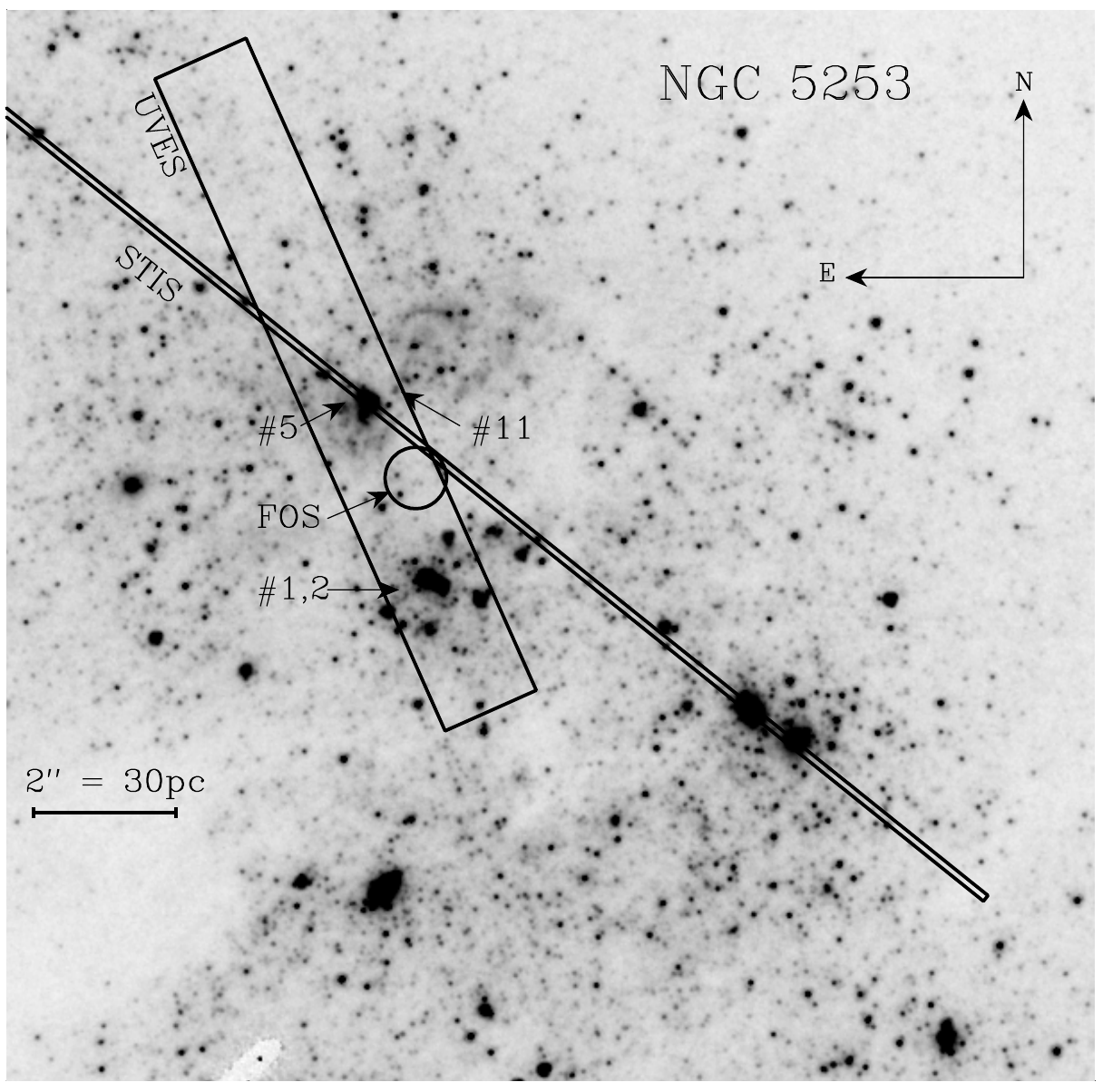}
\caption{HST ACS/HRC F550M image of the central region of NGC 5253 showing the positions of the HST/STIS and VLT/UVES slits, and the HST/FOS aperture. The two nuclear star clusters \#5 and \#11 \citep{calzetti15} are indicated, as well as clusters \#1,2 corresponding to UV-1 \citep{meurer95}.
}
\label{fig-V-slit}
\end{figure}
 \begin{figure}
\centering
\includegraphics[width=0.45\textwidth,angle=0,clip=true]
{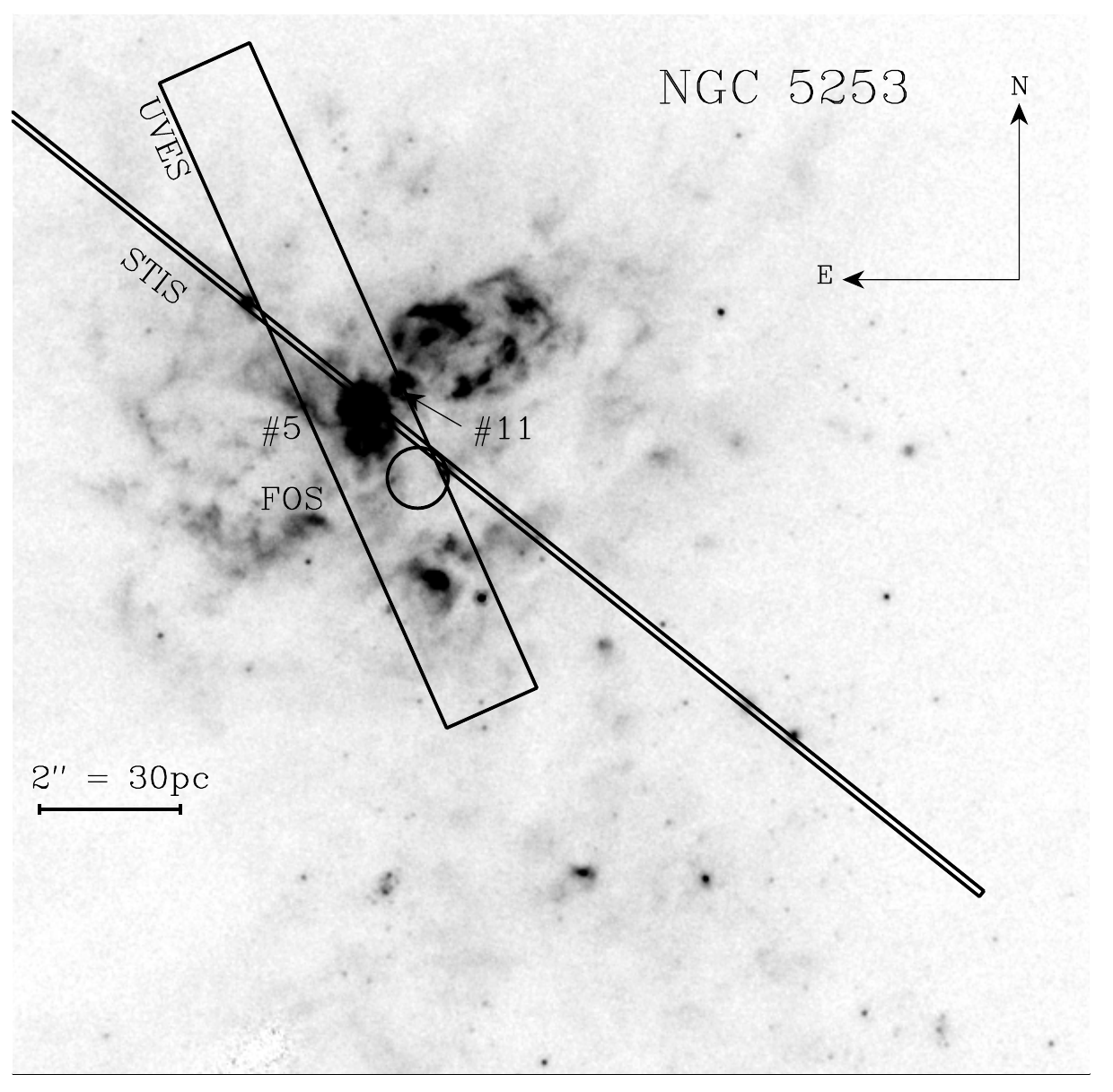}
\caption{HST ACS/HRC F658N image of the central region of NGC 5253 showing the complex structure of the ionized gas and the positions of the HST/STIS and VLT/UVES slits, the HST/FOS aperture, and clusters \#5 and \#11.}
\label{fig-Ha-slit}
\end{figure}
\begin{deluxetable*}{lllrccccl}
\tablecolumns{9}
%\rotate
\tabletypesize{\footnotesize}
\tablecaption{Emission and Absorption Line Measurements for NGC\,5253 Cluster \#5.\label{tab1}}
%\tablewidth{0pt}
\tablehead{
\colhead{Ion} & \colhead{Wavelength} & \colhead{Line Type} & \colhead{Velocity} & \colhead{FWHM} & \colhead{FWHM} & \colhead{$W_\lambda$} & \colhead{$10^{15}$\,Flux} & \colhead{Instrument}
\\
\colhead{} & \colhead{\AA} && \colhead{\kms} & \colhead{\AA} & \colhead {\kms} & \colhead{\AA} & \colhead{erg\,s$^{-1}$\,cm$^{-2}$\,\AA$^{-1}$} 
\\
}
\startdata
N\five\ & 1238.82, 1242.80  & Stellar P Cyg em. & $+$730 & \nodata & \nodata & $2.2\pm0.2$ & \nodata & STIS\\
O\five\ & 1371.30 & Stellar wind abs.& $-850$ & \nodata & \nodata & $1.0\pm0.2$ & \nodata & STIS\\
C\four\ & 1548.20, 1550.78 & Stellar P Cyg em. & $+$755 & \nodata & \nodata & $2.7\pm0.3$ & \nodata & STIS \\
&& Wind terminal vel. & $-2900$ & \nodata  & \nodata & \nodata & \nodata & STIS\\
He\two\ & 1640.42 & Stellar em.  & $+460$&$3.8\pm0.5$& $700\pm90$ &
$3.3\pm0.2$ & 1.8 & STIS\\
He\two\ & 1640.42 & Stellar em.  & $+140$&$7.0\pm1.2$ & $1270\pm210$ 
&$5.1\pm0.6$ & 3.7 & FOS\\
He\two\ & 1640.42 & Stellar em.  & $+270$&$6.5\pm0.6$ & $1180\pm100$
&$4.2\pm0.3$ & 2.5 & STIS$+$FOS\\
O\three] & 1660.81& Neb. em. &$-15$& $2.0\pm0.5$ & $370\pm90$
&$1.2\pm0.1$ &0.7 & STIS\\
O\three] & 1666.15 & Neb. em. &$-40$&$2.0\pm0.3$ & $360\pm50$
&$1.8\pm0.1$ & 1.2& STIS\\
C\three] & 1906.68, 1908.73 & Neb em. &$+25$ &$3.8\pm0.2$ & $590\pm30$
&$7.7\pm0.4$&3.7 & FOS\\
He\two\ & 4685.71  &Stellar em. & $+140$&$22.6\pm2.7$ & $1450\pm170$
&$1.8\pm0.1$ &0.380 & UVES\\
He\two\ & 4685.71  &Neb em. & $-17$ & $2.0\pm0.8$ & $130\pm50$
&$0.17\pm0.05$ & 0.063& UVES\\
\enddata
 \end{deluxetable*}

\section{Observations}\label{obs}
We have obtained high spectral resolution spectroscopy across the starburst core of NGC 5253 with the UV-Visual Echelle Spectrograph (UVES) at the Very Large Telescope (VLT) in Chile (Proposal ID 73.B-0238A, L.~J.~Smith, P.~I.).
We supplement this dataset with archival HST imaging obtained with the High Resolution Channel (HRC) of the Advanced Camera for Surveys (ACS) (Proposal ID 10609; W. Vacca, P. I.)
and far-ultraviolet (FUV) spectroscopy obtained with the Space Telescope Imaging Spectrograph (STIS) and the Faint Object Spectrograph (FOS). The STIS long-slit spectrum crosses cluster \#5 (Proposal ID 8232; D. Calzetti, P. I.) and has been presented in \citet{tremonti01} and \citet{chandar04}. The FOS spectrum (Proposal ID 6021; H. Kobulnicky, P.I.) covers a position close to cluster \#5 and the data are presented in \citet{kobulnicky97}.
All the HST archival data were obtained from the Mikulski Archive for Space Telescopes (MAST) and re-processed using the standard data-reduction pipelines.

In Figs.~\ref{fig-V-slit} and \ref{fig-Ha-slit}, we show the positions of the UVES, STIS and FOS observations superimposed on ACS/HRC F550M and F658N images of the central region of NGC 5253.

\subsection {STIS Spectra}
Long-slit UV spectra of the nuclear region of NGC 5253 were obtained with HST/STIS on 1999 July 27 and comprise 4 exposures (totalling 10235 s) taken with the G140L grating (1150--1730~\AA) and the $52''\times0''.1$ slit. The slit position is shown in Fig.~\ref{fig-V-slit} and covers $23''$ on the detector. The image scale is $0.0246$ arcsec/pixel ($=0.4$ pc/pixel) in the spatial dimension and 0.58 \AA/pixel in the spectral dimension. The spectra were co-added and extracted using a 22 pixel ($=0''.54=8.3$ pc) wide aperture for cluster \#5 and the sky regions defined in \citet{tremonti01}. The STIS spectral resolution is $\approx 1.8$ \AA\ or $\sim 360$ \kms\ \citep{tremonti01} and the spectrum was smoothed using $\sigma=0.4$ \AA, corrected for the radial velocity of this region (395 \kms; see next section) and binned to a pixel size of 0.75 \AA. The nebular O\three] $\lambda\lambda 1661, 1666$ lines have measured Gaussian widths of 2.0~\AA\ (Table~\ref{tab1}), in agreement with the expected spectral resolution.

\subsection{FOS Spectra} 
The FOS spectra were obtained on 1997 March 13 for a region close to cluster \#5 designated HII-1 with the G190H grating (1590--2310~\AA). Two exposures were taken of 2400 and 1240~s using the $0''.86$ circular aperture. The spectrum is presented in \citet{leitherer11} and we use their observed spectrum (corrected for the radial velocity) in preference to that available from MAST because it has a better wavelength calibration.
 The spectral resolution is 1.4~\AA\ for a point source and 3.2~\AA\ for a uniform source. The nebular C\three] emission line doublet has a FWHM of 3.8~\AA\ (Table~\ref{tab1}) or 3.1~\AA, when corrected for the wavelength separation of the doublet, in good agreement with that expected for a uniform source.
 FOS optical spectra are also available but the resolution and signal-to-noise is too low for them to be useful compared to the ground-based optical spectra we describe in the next section.
 
 \subsection{UVES Spectra}
The echelle spectra were obtained in service mode on 2004 May 4 with UVES at the VLT.
At the time of observation, the red arm of UVES contained a mosaic of an EEV and a MIT-LL CCD. The blue arm had a single
EEV CCD; all three CCDs have a pixel
size of 15~$\mu$m.  Simultaneous observations in the blue and the red
were made using the standard setups with dichroic \#1 (346+564~nm) and
dichroic \#2 (437+860~nm), covering an almost continuous wavelength
region from 3100--10360{\AA}; the regions between 5610--5670{\AA} and
8540--8650{\AA} were not observed as a result of the gap between the
two CCDs in the red arm.

\begin{figure*}
\centering
\includegraphics[width=12cm,angle=-90,clip=true]
{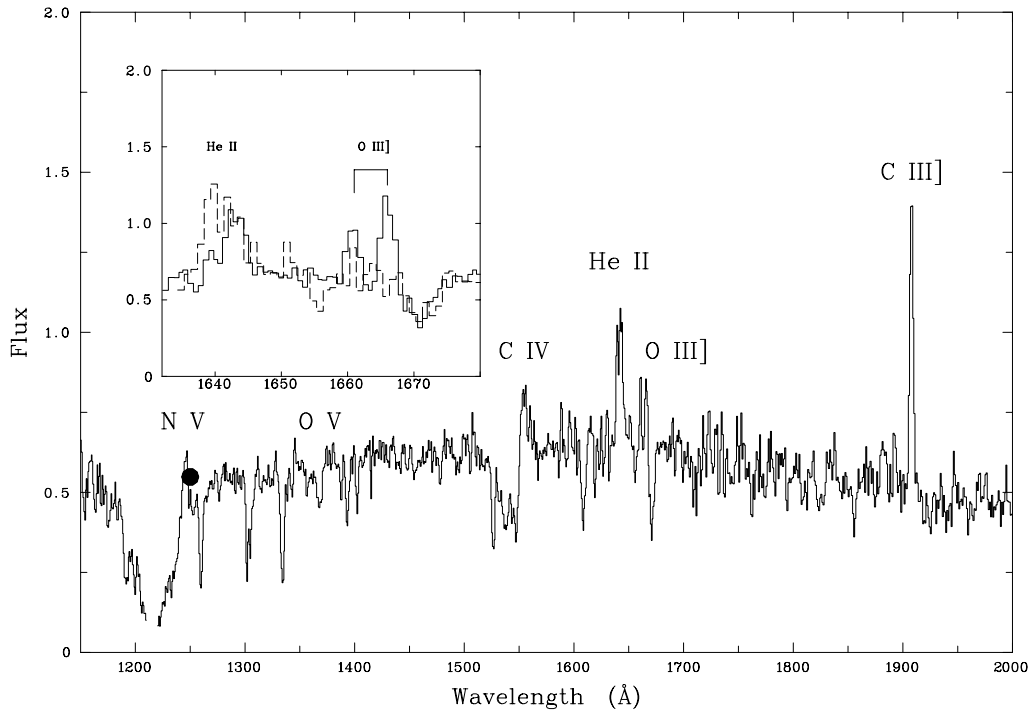}
\vskip-1.0truecm
\caption{Merged STIS and FOS spectrum of cluster \#5 over the wavelength range 1150--2000~\AA. The inset plot shows the overlap region covering the He\two\ $\lambda 1640$ emission feature for the STIS (solid line) and FOS (dashed line) spectra. The main stellar features and nebular emission lines are identified. The filled circle represents the measured luminosity density in the F125LP filter for cluster \#5 from \citet{calzetti15}. The flux is in units of $10^{-15}$\, ergs\,cm$^{-2}$\,s$^{-1}$\,\AA$^{-1}$.
}
\label{fig-UVspec}
\end{figure*}

The slit was chosen to have a position angle of 24{\degr}. As shown in Fig.~\ref{fig-V-slit}, it passes through cluster \#5 and cluster \#11 is on the edge of the slit. Clusters \#1 and 2 \citep{calzetti15} lie 2.5 arcsec south-west of cluster \#5 and correspond to the peak of the UV emission in NGC 5253 and are denoted UV-1 by \citet{meurer95}.
The slit width was 1\farcs4, giving a resolving power of $\sim 30\,000$ in the blue
and $\sim 28\,000$ in the red. 

For the observations made with dichroic \#1, the
target was observed at airmasses ranging from 1.5--1.9 with the seeing
typically varying between 0.8--0.9~arcseconds. For dichroic \#2, the range in airmass was 1.2--1.4 and the seeing
was 1.0--1.4 arcseconds. In total, three
exposures were taken for each waveband: two of 1425~s and one of 60~s.
The third, shorter exposure frames were taken to avoid saturating the 
strongest emission lines (e.g., the [O\three] lines and {\ha}). The slit length varied between 10 and 12 arcsec depending on wavelength.

The observations were reduced using the Common Pipeline Library version of the ESO UVES pipeline.
Spectra were extracted from the two-dimensional spectra for cluster \#5 (6 pixels or 3\farcs3; 1 pixel $=$ 0\farcs55) and also clusters \#1,2 (5 pixels or 2\farcs8) for comparison purposes. Sky subtraction was performed using pixels at the northern end of the slit.
The full spectrum for all clusters detected and a nebular line analysis has been presented by \citet{sidoli10}.

\section{Description of the Spectra}
\subsection{STIS and FOS UV Spectra}
In Fig.~\ref{fig-UVspec}, we show the STIS and FOS UV spectra over the spectral range 1150--2000~\AA. We plot them together even though they were acquired at different positions (see Fig.~\ref{fig-V-slit}) for the following reasons. There is a small overlap between the spectra in the region of the broad He\two\ emission (see inset in  Fig.~\ref{fig-UVspec}) and they both show this feature, although the profiles are slightly different, as we discuss below. To match the continuum levels, the STIS spectrum has been multiplied by a factor of 1.5. The observed F125LP filter luminosity density for cluster \#5 from \citet{calzetti15} is plotted in Fig.~\ref{fig-UVspec} and is in good agreement with the corrected STIS flux level. 

\begin{figure*}
\centering
\includegraphics[width=12cm,angle=-90,clip=true]
{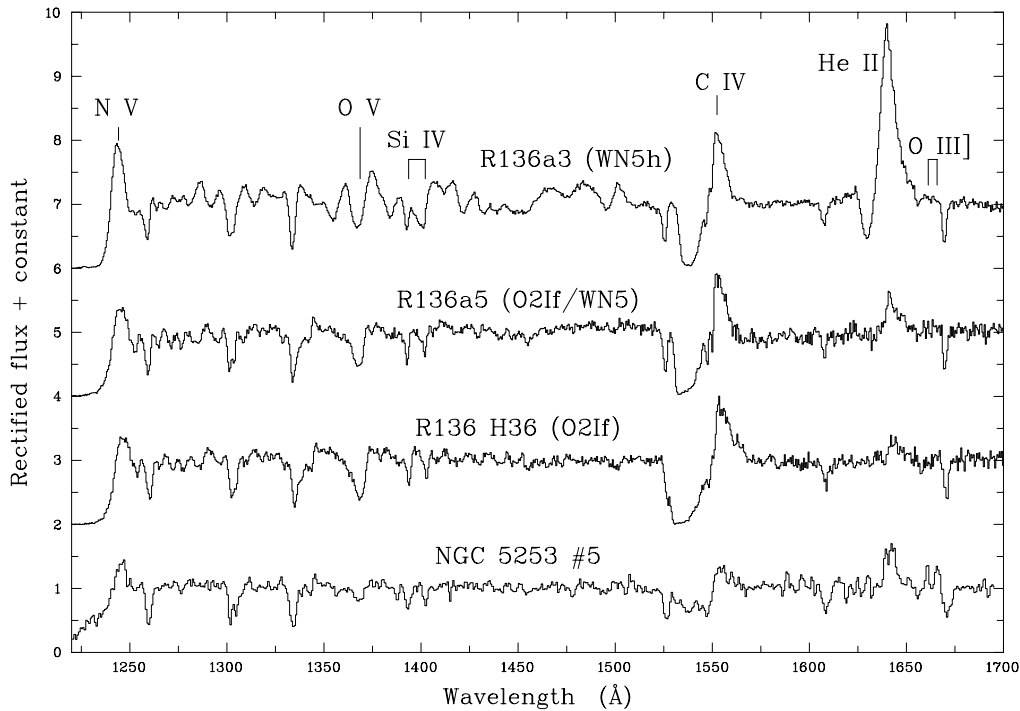}
\vskip-1.0truecm
\caption{Rectifed HST/STIS$+$FOS spectrum of cluster \#5 compared with HST/STIS spectra of R136a3, R136a5 and R136 H36 \citep{crowther16}.  The main spectral features are identified.}
\label{fig-UV-rect-spec}
\end{figure*}

The fact that the FOS spectrum has a strong continuum and broad He\two\ $\lambda1640$ leads us to doubt that it represents a pure nebular spectrum offset from cluster \#5. Yet, the spectrum was taken by offsetting from a star 40 arcsec to the SW, which was acquired correctly. The position of the FOS aperture shown in Fig.~\ref{fig-V-slit} was located using the offsets of \citet{kobulnicky97} and the position of the offset star on an ACS/F814W image of NGC 5253. Thus we believe the FOS aperture location is accurate and not centered on cluster \#5. We suspect that the FOS spectrum represents reflected and/or scattered light from cluster \#5 and/or cluster \#11. We consider this further in Sect.~4. For now, we assume that the STIS and FOS spectra both represent cluster \#5 and merge the two spectra in the overlap region. We note that we wish to compare UV and optical stellar emission features and the spatial extent of the UVES spectrum covers both cluster \#5 and the FOS aperture location.

In Fig.~\ref{fig-UV-rect-spec}, we compare the cluster \#5 UV spectrum with G140L STIS spectra of massive stars in the 1.5 Myr old LMC cluster R136 \citep[ID 12465]{crowther16}. The stars plotted are R136a3 (WN5h; $M = 180\pm30$\,\Msun), R136a5 (O2If/WN5;  $M = 100\pm30$\,\Msun) and  R136 H36 (O2If; $M = 70\pm10$\,\Msun), with current mass values taken from \citet{crowther16}. These stars have been chosen because their spectral types bracket the spectral characteristics of the cluster \#5 spectrum.

The similarity in the spectral features is striking, although the emission features are weaker in cluster \#5 due to the extra continuum contribution from other cluster O stars in the spectrum. The emission and absorption line measurements for cluster \#5 are presented in Table~\ref{tab1}. N\five\ $\lambda1240$, C\four\ $\lambda 1549$ and He\two\ $\lambda 1640$ are present in emission. 
O\five\ $\lambda1371$ is detected in cluster \#5 as a blueshifted absorption component with a velocity of $-850$~\kms\ (Table~\ref{tab1}), indicating a wind-affected O\five\ absorption. 
The terminal velocity measured from the C\four\ P Cygni absorption component is $-2900$~\kms, where the edge velocity has been corrected for turbulent motions in the outflow, following \citet{crowther16}.
All four spectra have weak to absent Si\four\ $\lambda 1393, 1402$, with interstellar absorption providing the dominant contribution. Similarly, with the exception of R136a3, they have negligible N\four] $\lambda1486$ emission.

\begin{figure*}
\centering
\includegraphics[width=12cm,angle=-90,clip=true]
{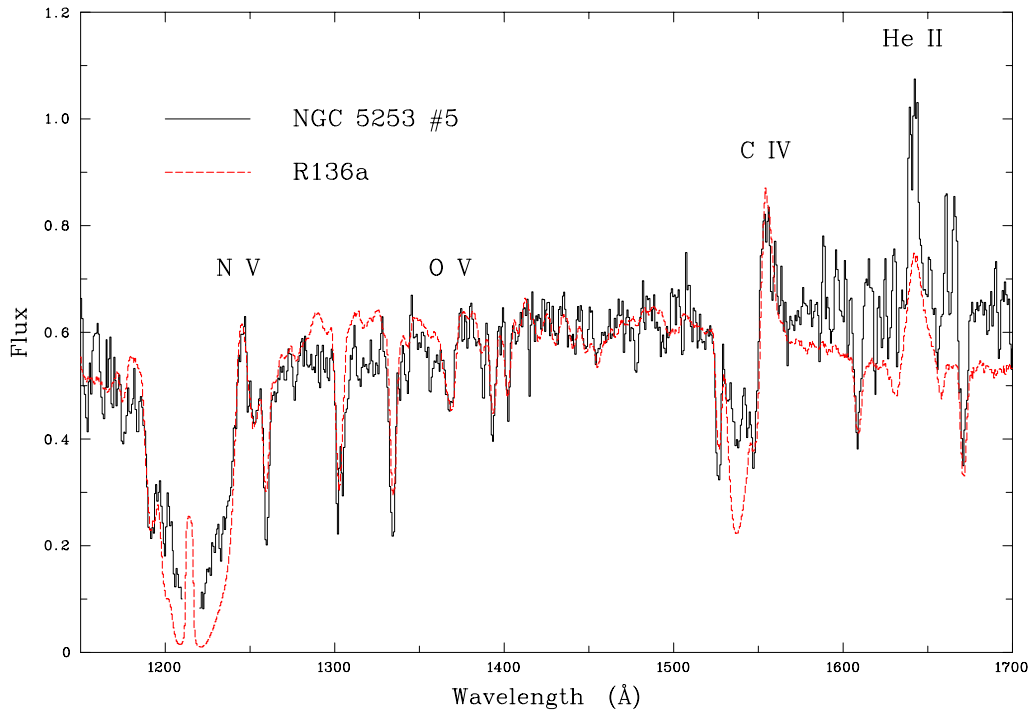}
\vskip-1.0truecm
\caption{Integrated HST/STIS spectrum of R136a \citep{crowther16} compared with the HST/STIS$+$FOS spectrum of cluster \#5. The R136a spectrum has been scaled to the distance of NGC 5253 and multiplied by 0.92.}
\label{fig-r136a_comp}
\end{figure*}

We now discuss the He\two\ $\lambda1640$ emission, which is present in both the STIS and FOS spectra of cluster \#5. This emission feature is resolved in both spectra and we provide measurements in Table~\ref{tab1}. The feature is broader and stronger in the FOS spectrum compared to the STIS spectrum (Fig.~\ref{fig-UVspec}). Beyond 1600~\AA, the sensitivity of the STIS G140L grating drops by a factor of 7 and the spectrum becomes very noisy. We adopt a He\two\ $\lambda1640$ FWHM of 1200~\kms\ from the merged STIS and FOS spectra. For comparison, the FWHMs of He\two\ $\lambda1640$ in the R136 stars are 1490 (a3), 1040 (a5) and 1100 \kms\ (H36).

Comparison of the relative strengths of the C\four\ and He\two\ emission features in Fig.~\ref{fig-UV-rect-spec} shows that the cluster \#5 spectrum is most similar to the WN5h star R136a3 and the O2If/WN5 star R136a5. The lack of Si\four\ P Cygni profiles and negligible or weak N\four] $\lambda1486$ also indicates a WN5 or O2~If/WN5 spectral type \citep{crowther98b}. 

We now compare in Fig.~\ref{fig-r136a_comp} the observed UV spectrum of cluster \#5 with the integrated STIS spectrum of R136a from \citet{crowther16}. To make this comparison, we simply scaled the R136a fluxes to the distance of NGC 5253 and tweaked the fluxes by multiplying the scaled R136a spectrum by 0.92 to match the central part of the cluster \# 5 spectrum. The agreement is excellent by just applying this scaling. We did not correct for the different reddenings because the color excesses of \#5 and R136a are similar with values of 0.46 \citep{calzetti15} and 0.42  \citep{crowther16}. 

The strengths, widths and velocities of the N\five, O\five\ and C\four\ lines are in excellent agreement between the two spectra. The absence of Si\four\ P Cygni emission is also evident in both spectra.
The continuum longward of C\four\ is weaker in the R136a spectrum and the He\two\ emission FWHM at 1900~\kms\ is broader than the cluster \#5 value of 1200~\kms. \citet{crowther16} classify the R136a integrated spectrum as Of/WN on the basis of a He\two\ $\lambda1640$ equivalent width of 4.5~\AA, which is below the dividing line between early O supergiants and WN5 stars at 5~\AA\ \citep{crowther98b}.
The measured equivalent width for cluster \#5 is 4.2~\AA, also suggesting an Of/WN classification.

\citet{crowther16} find that 95\% of the He\two\ $\lambda1640$ emission flux in the R136a integrated spectrum originates in six stars with masses above 100~\Msun. 
They suggest that the integrated spectra of young star clusters that display prominent He\two\ $\lambda1640$ emission and  O\five\ $\lambda1371$ absorption without strong Si\four\  $\lambda\lambda1393$--1402 P Cygni emission are indicative of a mass function that extends beyond 100~\Msun, and a very young age of less than 2~Myr. For cluster \#5, the presence of a wind-affected O\five\ absorption gives an upper limit to the age of 2~Myr. This implies that the He\two\ $\lambda1640$ emission has to arise from very massive Of/WN or H-burning WN stars. The lack of prominent Si\four\ emission indicates that the He\two\ $\lambda1640$ emission cannot originate in classical He-burning W-R stars with ages of 3--4~Myr.

We interpret the striking similarity between the R136a and cluster \#5 spectra displayed
in Fig.~\ref{fig-r136a_comp} as showing that cluster \#5 contains VMS. The only other known example is the cluster NGC 3125-A1 with strong (EW$=7$~\AA) and broad (FWHM$=1400$~\kms) He\two\ emission 
and O~\five\ absorption \citep{chandar04,hc06,wofford14}.

\subsection{UVES Optical Spectrum}
The optical spectrum of cluster \#5 is dominated by nebular emission lines. The only discernible stellar features are underlying Balmer absorptions to the upper Balmer nebular emission lines, and broad He\two\ $\lambda4686$ emission. The Balmer continuum is strong, signifying a very young stellar population. \citet{calzetti15} find that strong Balmer nebular continuum emission is present in the HST/WFC3 F336W images in the region of cluster \#5 and \#11, and that nebular line emission is present in all the other optical bands.
The average velocity of the nebular emission lines is measured to be $394.6\pm1.7$\,\kms and we adopt a value of 395\,\kms.

The presence of W-R stars in the optical spectra of extragalactic star-forming regions is usually inferred from the presence of broad emission bumps in the continuum near $\lambda4650$ and $\lambda5808$. The stronger `blue bump' is due to N\three\ $\lambda$4634--4641 and He\two\ $\lambda4686$ emission associated with WN stars, and the weaker `red bump' is due to C\four\ $\lambda\lambda5801, 5812$ emission produced by WC stars. If WC stars are present, they can also contribute to the blue bump via C\three\ $\lambda4650$.
\citet{monreal10} map the blue bump in the central region of NGC 5253 with integral field spectroscopy (IFS). They find WN stars associated with several clusters  including clusters \#5 and \#1,2 (Fig.~\ref{fig-V-slit}).   Similarly, \citet{west13}  use IFS to map the W-R red bump. They do not detect this WC feature in cluster \#5 but find it is present in \#1,2.

\begin{figure}
\centering
\includegraphics[width=0.55\textwidth,angle=0,clip=true]{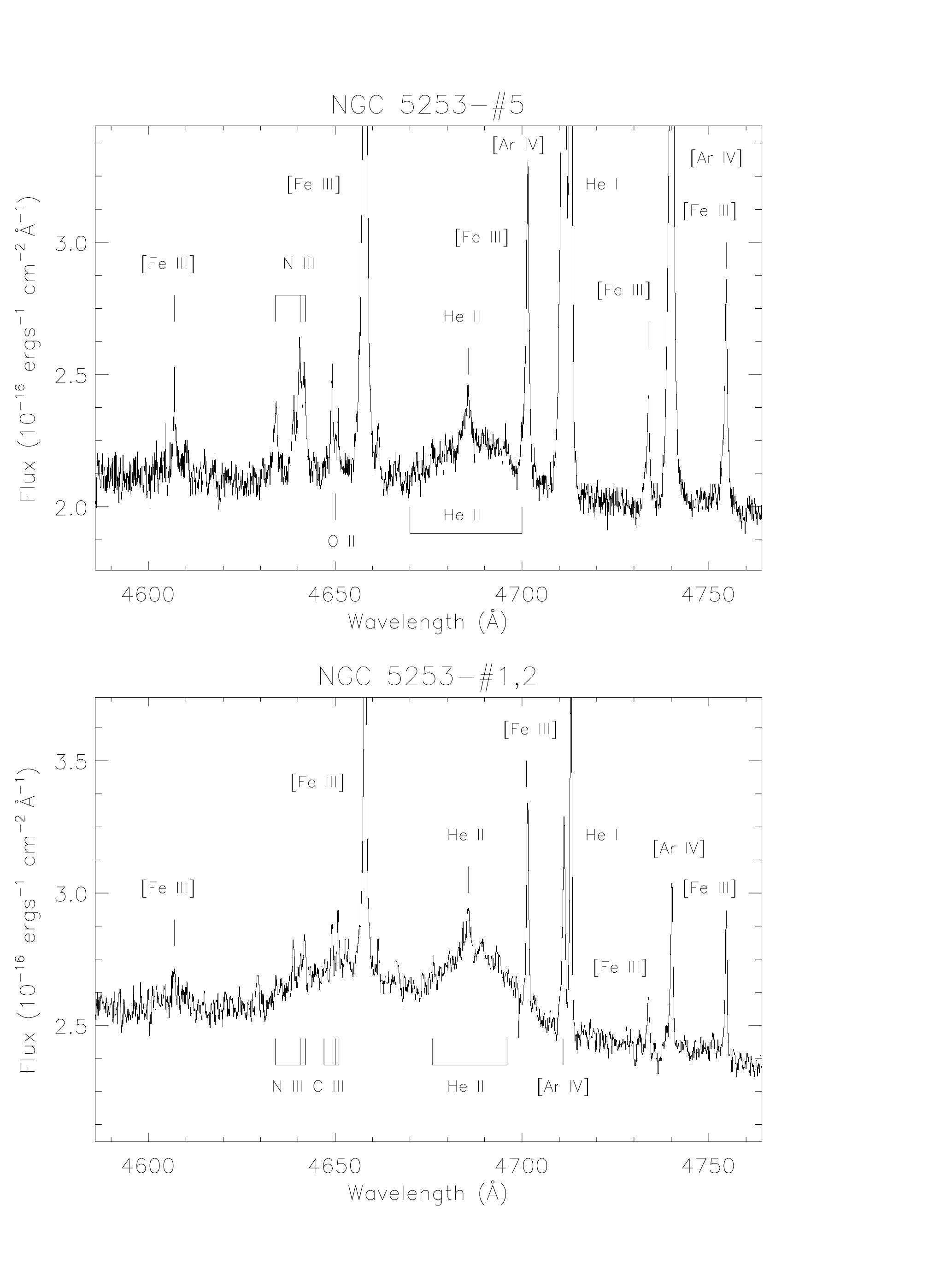}
\caption{VLT/UVES spectrum of cluster \#5 compared with clusters \#1,2. The main spectral features are identified.
}
\label{fig-WRbump}
\end{figure}

In Fig.~\ref{fig-WRbump}, we show the region of the spectrum covering the W-R emission features of 
N\three\ $\lambda3634$--4641, C\three\ $\lambda4647$--4651 and He\two\ $\lambda4686$ for clusters \#5 and \#1,2. Clusters \#1,2 show the classic features of the W-R blue bump arising from both WN and WC stars with N\three, C\three\ and He\two\ all present with broad emission profiles.
In contrast, cluster \#5 has broad He\two\ emission only. Both spectra show the presence of nebular He\two\ $\lambda4686$ emission.
N\three\ is present in the cluster \#5 spectrum but is narrow and the individual line components are resolved. The measured average line width of the three N\three\ lines is 1.3~\AA\ or 81~\kms\ compared to the measured width of 1.2\AA\ or 76~\kms\  for nearby nebular emission lines. The N\three\  lines are therefore likely to be nebular in origin, although detecting these transitions is unusual. 

The He\two\ $\lambda4686$ line measurements for cluster \#5 are given in Table~\ref{tab1}, corrected for nebular emission. The ratio of the nebular to stellar line flux is 17 per cent. The UVES fluxes in Table~\ref{tab1} have been scaled by 0.2 to match the photometry of \citet{calzetti15}. It should be noted that the extraction widths of the FOS ($0''.86$) and UVES ($3''.3$) apertures are very different and the fluxes of the W-R He\two\ emission features should not be directly compared because they sample different regions.

\begin{figure}
\centering
\includegraphics[width=0.5\textwidth,angle=0,clip=true]{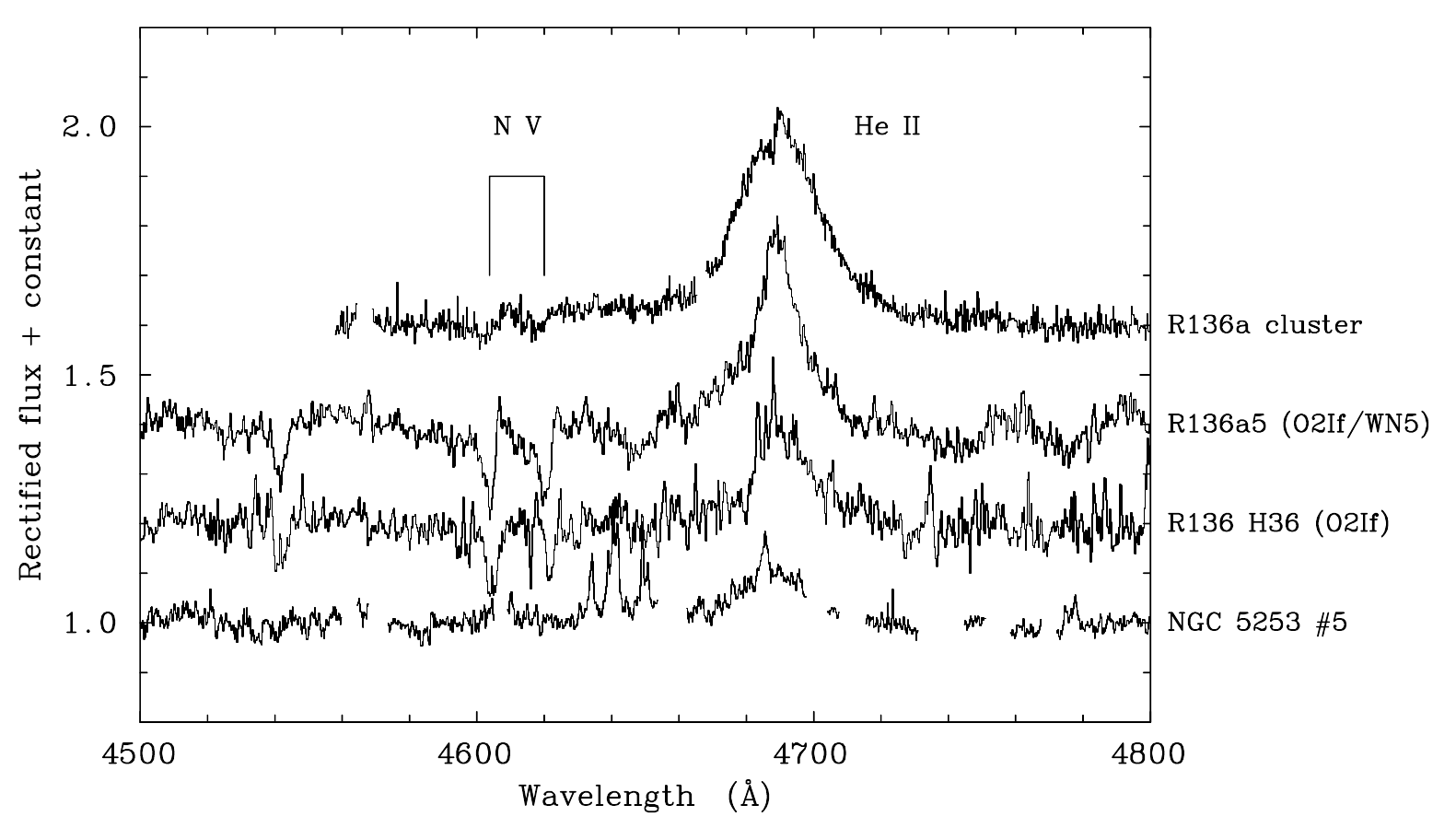}

\caption{Rectified UVES spectrum of cluster \#5 compared with HST/STIS spectra of R136a, R136a5 and R136 H36 (Caballero-Nieves et al., in prep.) in the region of the He\two\ $\lambda4686$ emission feature. The nebular features in the cluster \#5 spectrum have been removed.
}
\label{fig-opt}
\end{figure}

The broad He\two\ $\lambda 4686$ emission feature in cluster \#5 has a FWHM of 23~\AA\ or 1450 \kms\ (cf. 1200~\kms\ for He\two\ $\lambda1640$, Sect. 3.1). In Fig.~\ref{fig-opt}, we compare the cluster \#5 spectrum in the region of He\two\ $\lambda 4686$ to the R136a integrated cluster spectrum, and the R136a5 and R136 H36 stellar spectra, obtained with the STIS G430M grating (Caballero-Nieves et al., in prep.). All of the spectra have similar characteristics with broad He\two\ $\lambda 4686$ emission (1910; R136a, 1130; R136a5 and 1000~\kms; H36). The He\two\ emission in the R136a spectrum is much stronger than that seen in cluster \#5, and in other extragalactic star-forming regions with W-R stars present. This is probably because the R136a spectrum is a sum of the spectra of the most massive stars in the region and is not diluted by the continua of lower mass stars, which will become more significant in the optical compared to the UV.

The comparison of the optical spectra of cluster \#5 with R136a in the region of He\two\ $\lambda4686$ confirms our finding from the previous section that the classification is Of/WN. We base this on the similar width of the He\two\ feature as well as the absence of broad N\three\  emission. We note that the He\two\ $\lambda4686$ emission has to be intrinsically very strong to be visible in a young massive cluster spectrum because of the continuum dilution from other stars in the slit. It is unlikely that any He\two\ emission would be detected if only O supergiants were responsible.

\section{Discussion}
In Sect. 3, we show that the spectral emission line features in the UV and optical spectra of cluster \#5 can be classified as Of/WN.
This is the same spectral type as the very massive young stars in R136 \citep{crowther10, crowther16}, and the UV spectrum of cluster \#5 and R136a are almost identical. VMS have very dense, optically thick winds and show W-R-like broad emission lines while they are on the main sequence and hydrogen-rich.  However, the presence of O\five\ $\lambda1371$ wind absorption and the lack of UV Si\four\ P Cygni emission and broad optical N\three\ emission are incompatible with classic He-burning W-R stars. As \citet{crowther16} discuss, the presence of O\five\ provides an upper age limit of 2 Myr for R136. This is in accord with the very young age for cluster \#5 of $1\pm1$~Myr derived by \citet{calzetti15}. 
We thus conclude that the UV and optical spectra of NGC 5253-\#5 show that the cluster is 1--2 Myr old and contains stars with masses greater than 100~\Msun.

In Sect.~3, we  presented spectra for nuclear cluster \#5 only. Cluster \#11 has a projected separation from cluster \#5 of $0''.3$--$0''.4$ (5~pc)  and  is more massive and extreme in its properties \citep{turner15}. 
\citet{calzetti15} find that both clusters are $1\pm1$~Myr old but \#11 is heavily reddened and very faint at optical wavelengths. \citet{turner04} discuss whether cluster \#5 is a reflection nebula generated by cluster \#11. But, as discussed by \citet{calzetti15}, the HST imaging shows that cluster \#5 has a morphology that is consistent with a young, compact star cluster; it is slightly resolved with a size of $\sim 1.2$~pc, symmetric and centrally concentrated. Thus we will assume that there are two nuclear clusters, which only differ in mass and all other properties are similar. 

In Sect.~3.1, we presented FOS and STIS UV spectra, which both showed He\two\  $\lambda1640$ broad emission, even though the FOS spectrum was obtained $\sim 1''$ from the cluster \# 5 pointing for the STIS observation.  The FOS spectrum clearly represents scattered light from cluster \#5 and/or cluster \#11. Given that the two clusters have the same age,  we would also expect \#11 to contain VMS.  Cluster \#11 is very dusty \citep{calzetti15} and thus the broad He\two\ emission may originate from cluster \#11. Given that cluster \#11 is still deeply embedded in its natal gas and dust, it may  be slightly younger than cluster \#5, with an age of $< 1$~Myr. Even if this was the case, it would still be expected to host a significant number of VMS given its higher mass, and the VMS main sequence lifetime of $\approx 2$~Myr.

\subsection{Ionizing Fluxes}
One of the problems with matching the observations of the NGC 5253 nuclear clusters to standard stellar population synthesis models is the lack of sufficient ionizing photons at ages of 3-5~Myr. \citet{turner04} obtain a total ionizing flux $Q$(H\one)$ = 7 \times 10^{52}$~s$^{-1}$  or the equivalent of 7000 O7 stars 
for the central $1''.2$~(18 pc) from the observed 7~mm free-free continuum emission. One-third of this flux is confined to the central 5~pc region containing clusters \#5 and 11. %and one-fifth to the parsec-sized supernebula core containing cluster \#11.
\citet{turner15} take the total ionizing flux, an age of 4 Myr, the virial mass limit for this region of $1.8\times 10^6$ \Msol, and compare to the predicted ionizing fluxes from  Starburst99 \citep{leitherer99}. They find that to match the observations, a top-heavy initial mass function (IMF) is required with a lower mass cutoff of 3~\Msol.

This discrepancy can be solved if more ionizing photons are produced by the cluster stars than predicted by the stellar population synthesis models. There are potentially two ways this can be done: the inclusion of rotating massive stars \citep{levesque12, leitherer14}; and/or the inclusion of very massive stars \citep{crowther10, doran13}. We consider each in turn.

\citet{leitherer14} present new Starburst99 models which include stellar evolutionary tracks with rotation from the Geneva group for \Zsol\ and $\frac{1}{7}$ \Zsol\ \citep{ekstrom12, georgy13}.
At the time, the available rotating tracks were limited and the authors discuss the extreme case when the massive stars are rotating at 40\% of their breakup velocity. 
Models with this level of rotation are both hotter because of the larger helium surface abundance and more luminous because of a larger convective core.
These two effects lead to a higher ionizing photon output in the Lyman continuum at ages $\ge 4$~Myr. 
However, measurements of the rotational velocities of single O-type stars in the 30 Dor region show that they rotate with a rate of less than 20\% of their breakup velocity \citep{ramirez13}. Thus the ionizing outputs of the \citet{leitherer14} models are likely to be over-estimates.

If we take the total of the derived masses for clusters \#5 and 11 of $3.3\times 10^{5}$~\Msol\ and the ionizing flux of $Q$(H\one)$ = 2.2 \times 10^{52}$~s$^{-1}$ for the central 5~pc region, Starburst99 models for a standard Kroupa IMF predict values of 1.3 (2 Myr) and 0.4 (4 Myr) $ \times 10^{52}$~s$^{-1}$ for the solar non-rotating case. For the fast-rotating models, these values rise to 1.5  $\times 10^{52}$~s$^{-1}$ for both the 2 and 4 Myr cases.
The metallicity of NGC 5253 is 35\% solar  \citep{monreal10} and $Q$(H\one) will increase compared to solar because of the lower wind opacities. Indeed, the non-rotating SMC metallicity tracks with Starburst99 predict values of $Q$(H\one) of 1.6 and 0.6 $ \times 10^{52}$~s$^{-1}$ for 2 and 4 Myr respectively. All of these values fall short of the required ionizing flux by 30--80\%.

If massive O stars rotate as fast as 40\% of their breakup velocity at SMC metallicity, the predicted values of $Q$(H\one) may increase sufficiently to agree with the observed ionizing flux. But this is unlikely to work because the comparisons above assume the cluster masses derived from the non-rotating Starburst99 models which fit the observed cluster SEDs from the UV to the IR \citep{calzetti15}. As \citet{leitherer14} discuss, the higher luminosities of fast rotating massive stars will lead to bluer SEDs and lower star cluster masses as the theoretical $L/M$ is higher. Thus it is unlikely that rotating massive stars can provide the deficit of ionizing photons in the nuclear clusters of NGC 5253.

\citet{doran13} provide a census of the hot luminous stars in R136 and the 30 Dor region of the LMC. The R136 cluster ($< 5$~pc) has a mass of $5.5\times 10^4$~\Msol\ \citep{hunter95} and an age of $1.5\pm0.5$~Myr \citep{crowther16}. It contains nine VMS \citep{crowther16}  inside the half-light radius of 1.7~pc \citep{hunter95}. \citet{doran13} find  $Q$(H\one)$= 7.5 \times 10^{51}$~s$^{-1}$ for the R136 region. 
They compare this output to non-rotating Starburst99 models and find that the predicted ionizing flux is a factor of two lower than their empirical value at 2~Myr. The upper mass in the Starburst99 models is 100~\Msol\ and thus the contributions from stars with masses higher than this are ignored. \citet{doran13} have eight such stars located in the R136 region and by excluding these stars, they find much better agreement with the Starburst99 models. The four WN5h stars in R136 have a combined ionizing flux of $Q$(H\one)$=  2 \times 10^{51}$~s$^{-1}$ and are responsible for 25\% of the ionizing flux. For the two NGC 5253 nuclear clusters, if we take the predicted ionizing flux from \citet{turner04} of $Q$(H\one)$ = 2.2 \times 10^{52}$~s$^{-1}$ and the predicted ionizing flux from Starburst99 of $Q$(H\one)$ = 1.6 \times 10^{52}$~s$^{-1}$ for SMC metallicity and non-rotating tracks at 2~Myr, then the deficit amounts to $6 \times 10^{51}$~s$^{-1}$ or the equivalent of 12 WN5h stars with masses $\ge 150$~\Msol. Given the six times greater mass for clusters \#5 and 11 compared to R136, this number seems reasonable. The addition of VMS will hardly change the cluster SED because of their small numbers. These types of stars are not expected to have an appreciable flux below the He\two\ Lyman limit because of their dense winds. Their presence cannot therefore explain the He\two\ $\lambda4686$ nebular emission (Sect. 3.2). 

\subsection{Origin of the Nitrogen Enrichment} 
\citet{monreal10, monreal12} use IFS to map out the nebular N enrichment. They find that it peaks ($\approx 3$ times overabundant) at the position of the two nuclear clusters  and covers an area $65 \times 30$ pc corresponding to the giant H\two\ region associated with the clusters (Fig. \ref{fig-Ha-slit}). The source of this enrichment is unexplained because the He abundance should also be enhanced if the pollution is caused by the chemically-enriched winds of W-R stars but this has not been seen with any certainty \citep{monreal13}. 

\citet{kohler15} present an evolutionary model grid at LMC metallicity for stars in the H-burning phase with initial masses from 70 to 500~\Msun\ and rotational velocities from 0 to 550~\kms. They find that N enrichment is ubiquitous because rotational mixing and/or mass loss quickly leads to the establishment of CNO-equilibrium abundances in the atmospheric layers. Whether surface He enrichment occurs is more complicated and depends on the mass of the star through the mass loss rate and the rotation velocity.

As an illustrative example, the \citet{kohler15} evolutionary track for a 100~\Msun\ star rotating at 200~\kms\ (20\% of the critical velocity) reaches CNO-equilibrium surface abundances at an age of 1~Myr, and the surface N abundance equals the enriched value observed in NGC 5253 at an age of only 0.26 Myr. In this model, the He abundance does not start to increase until an age of 2.2 Myr. At an age of 1~Myr, the star has lost 4.5~\Msun\ of its original mass. We now consider whether such massive, rotating O stars can produce the observed level of N enrichment over a timescale of $\sim 1$~Myr.

Observationally, the nebular emission line components are composed of three components with the N-enriched material confined to a broad component with a FWHM of 100--150~\kms,  an electron density of 500 \cmt, and a linear velocity gradient  from $-50$ to $+50$~\kms\ over the central 4$''$ centered on cluster \#5 \citep{sidoli10, monreal10,west13}. Taking the outflow velocity as 70~\kms\ and a size of 65~pc for the N-enriched region \citep{monreal12,west13}, we find a dynamical time scale for the pollution of 0.5~Myr. Thus there is sufficient time for clusters \#5 and 11 to have polluted their environment despite their young age of 1--2~Myr.

We can derive the mass of the excess N by assuming the region is cylindrical with a radius of 15~pc and an enriched $\log {\rm N/O} = -0.81$ compared to $-1.37$ in unenriched knot \#2 \citep{monreal12}. The resulting mass of excess N is 1--10~\Msol, assuming a volume filling factor of 0.01--0.1 for the ionized gas. If we assume an average stellar surface N enrichment of a factor of 10 \citep{kohler15}, giving a nitrogen mass fraction of $7.6\times 10^{-4}$, then the stellar mass loss rate  required to deposit 1~\Msol\ of N in 1 Myr is $1.3\times10^{-3}$~\Msol\ yr$^{-1}$. For a \citet{kroupa01} IMF extending to 300~\Msol, there should be 300 stars more massive than 50~\Msol\ for the combined cluster mass of $3.3\times10^5$~\Msol, giving a reasonable average stellar mass loss rate of $4.4\times 10^{-6}$ \Msol\ yr$^{-1}$. 
While these calculations are approximate, they demonstrate that the observed N enrichment associated with clusters \# 5 and 11 can be produced by the stellar winds of massive ($>50$~\Msun), rotating cluster O stars in the first million years. The presence of VMS in the two clusters will increase the amount of enriched nitrogen produced because they rapidly ($<0.5$~Myr) reach CNO equilibrium abundances on their surfaces and have very high mass loss rates ($>10^{-5}$ \Msol\ yr$^{-1}$), particularly if they are close to their Eddington limits.

At an age of 1~Myr, the lower mass stars in the NGC 5253 nuclear clusters will still be forming and thus may be nitrogen-enriched. This is of relevance to models seeking to explain multiple stellar generations in globular clusters using fast rotating massive stars 
as the source of self-enrichment \citep{decressin07}.

\subsection{Comparison with He\two-emitting High Redshift Galaxies}
The UV spectrum of cluster \#5 is representative of a very young, nearby, low metallicity, nuclear starburst. It bears a striking resemblance to the UV rest-frame spectrum of Q2343-BX418 \citep{erb10}.  These authors suggest this $z=2.3$ young, low mass, low metallicity galaxy is a plausible analog to the young, low metallicity star-forming galaxies at $z>5$.
Q2343-BX418 has broad He\two\ $\lambda1640$  emission (1000~\kms) with a rest equivalent width of 2.7~\AA\ and  nebular emission from O\three] $\lambda\lambda 1661, 1666$ and C\three] $\lambda\lambda 1907, 1909$ is also detected. The C\three] equivalent width of 7.1~\AA\ is very similar to the value of 7.7~\AA\ that we measure for the FOS spectrum near cluster \#5. \citet{rigby15} discuss measurements of C\three] emission in nearby and high redshift galaxies. In their nearby sample, they include the FOS spectrum of NGC 5253 used here, and note that this galaxy is one of the 20\% of local galaxies that show strong C\three] emission.

\citet{cassata13} identified 39 He\two\ $\lambda1640$ emitting galaxies over a redshift range of 2--4.6 from deep survey data. They classify the He\two\ emitters into two different classes depending on the width of the emission. They explain the broad emission (FWHM $> 1200$~\kms) with W-R stars and suggest that for the 11 objects with unresolved He\two\ $\lambda1640$ emission (FWHM $< 1200$~\kms), the emission is nebular in origin and either powered by the strong ionizing radiation field from a stellar population that is rare at $z \sim 0$, or Population III star formation. 

\citet{grafener15} discuss the possibility that the narrow He\two\ emission arises from a population of VMS at low metallicity and show that the predicted He\two\ line strengths and widths are in line with those expected for a population of VMS in young super star clusters located in the galaxies of the \citet{cassata13} sample. We note that we measure a He\two\ $\lambda1640$ line width of 1200 \kms\ for cluster \#5 which is at the borderline between the two groups of He\two\ emitters identified by \citet{cassata13}. Moreover, we identify this emission as originating in VMS.

The presence of VMS in young, high redshift galaxies has important consequences for the ionizing photon budget and cosmic reionization. As \citet{doran13} and this paper (Sect.~4.1) have shown, VMS increase the number of ionizing photons by up to a factor of two for the first 1--3~Myr of a starburst compared to standard spectral synthesis models, which typically have upper mass cutoffs of 100~\Msun. Interestingly,  \citet{zastrow13} find evidence for the escape of ionizing photons in two out of seven nearby dwarf starbursts through the detection of ionization cones. These two galaxies are NGC 5253 and NGC 3125, which also has a UV spectrum indicative of VMS \citep{wofford14}.

The James Webb Space Telescope (JWST) will obtain rest frame UV spectra of  high redshift star-forming galaxies. These spectra may reveal the presence of VMS through the presence of He\two\ emission and O\five\ wind absorption, and the absence of Si\four\ P Cygni emission. It is crucial to extend the current suite of stellar population synthesis models to cutoff masses greater than 100~\Msun\ in order to model their properties correctly.

\section{Summary and Conclusions}

We have examined UV and optical spectroscopy of cluster \#5 in the nucleus of NGC 5253 with the aim of reconciling the extremely young age of $1\pm1$~Myr found by \citet{calzetti15} with the presence of W-R features in the cluster spectrum. Specifically, we have investigated whether the W-R features arise from hydrogen rich, very massive stars with masses greater than 100~\Msun. In addition, the presence of VMS may be necessary to account for the 50\% deficit in the ionizing photon rate \citep{calzetti15} and the nitrogen-enriched nebular gas.

We present archival STIS and FOS UV spectra of cluster \#5 and compare them with STIS spectra of VMS in R136, the central ionizing cluster of 30 Dor in the LMC, and the R136a integrated spectrum from \citet{crowther16}. We detect broad (1200~\kms) He\two\ $\lambda1640$ emission, O\five\ $\lambda1371$ wind absorption, C\four\ P Cygni emission and an absence of Si\four\ P Cygni emission. These spectral characteristics are only compatible with a very young age ($<$~2 Myr) and a mass function that extends beyond 100~\Msun. The UV spectrum of \#5 is extremely similar to the integrated spectrum of the R136a cluster, which contains six VMS \citep{crowther16}. 

We compare the cluster \#5 VLT/UVES spectrum in the region of He\two\ $\lambda4686$ to another cluster in NGC 5253, containing classical W-R stars, the individual VMS in R136 and the R136a cluster. We find that cluster \#5 has broad He\two\ emission and an absence of other broad features due to N\three\ or C\three, which are usually seen in clusters containing W-R stars.  Again, the spectrum is most similar to the R136a spectrum and confirms our finding from the UV spectrum that VMS are present in cluster \#5.

We conclude from the UV and optical comparisons that NGC 5253-5 is very young with an age of less than 2~Myr and that the broad He\two\ emission is produced by VMS that have dense, optically thick winds and show W-R-like emission features while they are on the main sequence.

We consider whether the 50\% discrepancy in the predicted and observed ionizing flux for the nuclear region of NGC 5253 can be solved by increasing the number of ionizing photons produced by the cluster stars. We compared $Q$(H\one) to the output of Starburst99 models that include fast rotating massive star evolutionary tracks \citep{leitherer14}. We find these models are unlikely to work because they do not produce sufficient ionizing photons and alter the cluster SED, which is well-fitted by non-rotating Starburst99 models \citep{calzetti15}.
We consider whether VMS can account for the photon deficit by using the empirical results of \citet{doran13} for R136. They find that the four WN5h stars account for 25\% of the R136 ionizing flux. At an age of 2 Myr, the predicted ionizing flux from Starburst99 \citep{leitherer99} at SMC metallicity is 30\% less than the observed ionizing flux within the central 5~pc. The equivalent of 12 WN5h with masses $> 150$~\Msun\ is required to make up the deficit, which is reasonable given that the combined cluster mass of  \#5 and \#11 is six times more massive than R136.

We investigated the origin of the nitrogen enrichment in the giant H\two\ region surrounding the nuclear clusters. The dynamical timescale for the enrichment is 0.5~Myr, well within the age of 1--2~Myr for the nuclear clusters. We find that the mass of enriched nitrogen is 1--10~\Msun\ depending on the filling factor of the ionized gas. Rapid surface nitrogen enrichment occurs through mass loss and/or rotational mixing
in the new evolutionary model grid of \citet{kohler15}, which includes stars up to 500~\Msun\ with rotation for LMC composition. We find that the mass of excess N can be produced by rotating massive stars in the clusters if surface CNO equilibrium abundances are achieved within the first 1 Myr, as indicated by the \citet{kohler15} models. We note that the lower mass stars, which are presumably still forming, may be nitrogen rich. The nuclear region of NGC 5253 is probably observationally unique in containing very young star clusters, which have already polluted their environments with the chemically-enriched winds of massive stars.

We compare the UV spectrum of cluster \#5 with those of metal poor, high redshift galaxies and show that it has many similarities in terms of the He\two\ emission line strength and width, and the presence of strong O\three] $\lambda\lambda 1661, 1666$ and C\three] $\lambda\lambda 1907, 1909$ nebular emission. VMS may exist in young star-forming regions at high redshift and their presence should be revealed by UV rest frame spectra to be obtained by JWST. Population synthesis models typically have upper mass cutoffs of 100~\Msun. It is crucial to extend these into the VMS regime to correctly account for the radiative, mechanical and chemical feedback, which will be dominated by VMS for the first 1--3~Myr in star-forming regions.

\acknowledgments

We thank Ed Churchwell, Jay Gallagher, Selma de Mink, Hugues Sana, Christy Tremonti and Aida Wofford for enlightening discussions.
Based on observations collected at the European Organisation for Astronomical Research in the Southern Hemisphere 
under ESO program 73.B-0238(A).
Based on observations made with the NASA/ESA Hubble Space Telescope, obtained from the Mikulski Archive for Space Telescopes (MAST) at the Space Telescope Science Institute, which is operated by the Association of Universities for Research in Astronomy, Inc., under NASA contract NAS5-26555. 

{\it Facilities:} \facility{HST (STIS, FOS, ACS/HRC}, \facility{VLT:Kueyen (UVES)}

%%%%%%%%%%%%%%%%%%%%%%%%%%%%%%%%%%
% Bibliography
%%%%%%%%%%%%%%%%%%%%%%%%%%%%%%%%%%
\bibliographystyle{apj}
\bibliography{references}

\end{document}